\documentclass[11pt]{article}
\usepackage{epsfig} 
\newcommand{\sect}[1]{ \section{#1} \setcounter{equation}{0} }

\newcommand{\pslash}{p \! \! \! /}

\newcommand{\half}{\mbox{\small{$\frac{1}{2}$}}} 
 
\newcommand{\Nf}{N_{\!f}} 
\newcommand{\MSbar}{\overline{\mbox{MS}}} 

\begin{document}
\title{Three loop DIS and transversity operator anomalous dimensions in the
RI$^\prime$ 
scheme\thanks{Talk presented at $11$th International Workshop on Advanced 
Computing and Analysis Techniques in Physics Research, NIKHEF, Amsterdam, The 
Netherlands, $23$rd-$27$th April, 2007.}
} 
\author{J.A. Gracey, \\ Theoretical Physics Division, \\ 
Department of Mathematical Sciences, \\ University of Liverpool, \\ P.O. Box 
147, \\ Liverpool, \\ L69 3BX, \\ United Kingdom.} 
\date{} 
\maketitle 
\vspace{5cm} 
\noindent 
{\bf Abstract.} We discuss the computation of the three loop anomalous
dimensions for various operators used in deep inelastic scattering in the 
$\MSbar$ and RI$^\prime$ schemes. In particular the results for the $n$~$=$~$5$
and $6$ Wilson operators in arbitrary linear covariant gauge in the RI$^\prime$
scheme are new.  

\vspace{-15.5cm}
\hspace{13.5cm}
{\bf LTH 747} 

\newpage 

\sect{Introduction.}

The use of symbolic manipulation and computer algebra has been an invaluable
tool for providing large loop order results in quantum field theory in general
and in quantum chromodynamics (QCD) in particular. The outstanding example is 
that of the full three loop anomalous dimensions for flavour non-singlet and
singlet unpolarized operators for deep inelastic scattering as an analytic
function of the operator moment $n$, \cite{1,2,3,4}. Also the Wilson 
coefficients have been provided to the same precision. This large project, of 
the order of ten years, required not only the extensive use of the symbolic
manipulation programme {\sc Form}, \cite{5}, but also its own development to 
handle the unforeseen complexity of the computation. An earlier approach, 
\cite{6,7}, to this problem in deep inelastic scattering was to determine 
results for fixed (even) moments using the {\sc Mincer} algorithm, \cite{8}, 
translated into {\sc Form}, \cite{9}. For instance, the first few moments were 
given in \cite{6,7} and subsequently those for $n$~$=$~$12$ and $n$~$=$~$16$ 
appeared in \cite{10,11}. At that time the fixed moment expressions were used 
to parametrize the full expressions but such an approach was clearly incomplete 
lacking the correctness of a full evaluation. However, the results subsequently
served as very important independent checks on the final arbitrary $n$ results.
Now that the computational algorithm has been established, in principle it can 
be applied to other operators underlying related phenomenology. For example, 
the case of polarized Wilson operators will be relevant for spin physics. In 
addition in the spin context there is interest in a similar operator called
transversity, \cite{12,13,14}. This corresponds to the probability of finding a
quark in a transversely polarized nucleon polarized parallel to the nucleon 
versus that of the nucleon in the antiparallel polarization. From a theoretical
point of view it is similar to the non-singlet unpolarized Wilson operator but 
experimentally it is not as accessible since there is no direct coupling to 
quarks. Nevertheless there have been proposals to study it at RHIC. Therefore, 
whilst in principle it is possible to calculate the arbitrary moment three loop 
transversity operator anomalous dimensions in the $\MSbar$ scheme, it would be 
important to have strong independent checks on any future full result. Akin to 
the 1990's approach for the Wilson operator there is therefore a need for fixed
moment calculations. Aside from this motivation, there is a secondary one. 

One of the ingredients necessary to study the structure functions is the
measurement of the non-perturbative matrix elements. From the theory point of
view, a tool which achieves this is lattice regularization and various groups, 
such as QCDSF, have developed a substantial programme to determine key matrix 
elements. (See, for instance, \cite{15,16,17}.) However, one technical aspect 
of such work is ensuring that the results agree in the continuum with 
expectations from the ultraviolet limit. One approach in this respect is for 
the lattice results to be matched onto the perturbative expressions in the 
chiral limit, where to aid precision one would prefer the results to as high a 
loop order as is calculationally feasible. This has been considered in a series
of articles, \cite{18,19,20}, to three loops. One technical issue is that to 
keep time (and money) to a minimum, the lattice computations are performed in 
renormalization schemes known as regularization invariant (RI) and its 
modification, RI$^\prime$, \cite{21,22}. Unlike the $\MSbar$ scheme, they are 
mass dependent renormalization schemes. Results in this scheme have then to be 
converted to the standard $\MSbar$ scheme. In the continuum QCD has been 
renormalized at high loop order in both RI and RI$^\prime$ in \cite{23,18} and 
the conversion functions established for various quantities of interest. 
Therefore to aid lattice computations of the matrix elements one requires the 
finite part of the analogous Green's functions but only for low moment since 
clear signals for higher moments information are hard to extract from the 
numerical noise on the lattice. Given the need for such accurate results for a 
specific Green's function on the lattice, not only for the transversity 
operators but also for the Wilson operators, we report the results of recent 
computations in this area, \cite{20,24}. One consequence of the Green's 
function considered is that the operator anomalous dimensions emerge as a 
corollary in the RI$^\prime$ and $\MSbar$ schemes. In addition to the results 
already available, \cite{18,19,20,24}, we will provide the RI$^\prime$ 
anomalous dimensions for the $n$~$=$~$5$ and $6$ Wilson operators in arbitrary 
covariant gauge. Although the finite parts are required for lattice 
calculations for low moment, our computations have been extended to high moment
for the transversity operator but only to determine the anomalous dimensions. 
At the appropriate point we will mention some of the computer algebra aspects 
of this and previous work which without the power of {\sc Form} would have 
rendered the determination of any result virtually impossible.

The paper is organised as follows. In section two we introduce our notation and
computational strategy in more detail before discussing the appropriate points
of the RI$^\prime$ scheme in section three. Section four details a simple low
moment example, whilst the symbolic manipulation issues are recorded in the
context of the higher moment calculation in section five. The explicit 
anomalous dimensions are given in section six with a few concluding remarks
provided in the final section. 

\sect{Background.}

The two basic classes of operators we will consider are the non-singlet Wilson 
operators  
\begin{equation}
{\cal O}_W^{\nu_1 \ldots \nu_n} ~=~ {\cal S} \bar{\psi}^i \gamma^{\nu_1} 
D^{\nu_2} \ldots D^{\nu_n} \psi^j 
\label{wilop}
\end{equation} 
and the transversity operators 
\begin{equation}
{\cal O}_T^{\mu \nu_1 \ldots \nu_n} ~=~ {\cal S} \bar{\psi}^i \sigma^{\mu\nu_1} 
D^{\nu_2} \ldots D^{\nu_n} \psi^j 
\label{traop}
\end{equation}
where $\psi^i$ is the quark field, $1$~$\leq$~$i$~$\leq$~$\Nf$ for $\Nf$ quark
flavours, $D_\mu$ is the covariant derivative and
$\sigma^{\mu\nu}$~$=$~$\half [ \gamma^\mu, \gamma^\nu ]$. The operation of
totally symmetrizing with respect to the Lorentz indices $\{\nu_i\}$ and 
ensuring the operator is traceless is denoted by ${\cal S}$ where the
respective but different tracelessness conditions are given by  
\begin{equation}
\eta_{\nu_i\nu_j}{\cal O}_W^{\nu_1\ldots\nu_i\ldots\nu_j\ldots\nu_n} ~=~ 0 
\end{equation}
and, \cite{25},  
\begin{equation}
\eta_{\mu\nu_i}{\cal O}_T^{\mu\nu_1\ldots\nu_i\ldots\nu_n} ~=~ 0 ~~~~ 
(i ~\geq~ 2) ~~~,~~~ 
\eta_{\nu_i\nu_j}{\cal O}_T^{\mu\nu_1\ldots\nu_i\ldots\nu_j\ldots\nu_n} 
~=~ 0 ~. 
\label{trasymm} 
\end{equation}
For (\ref{traop}) the anomalous dimensions are available for all moments $n$ at
one and two loops in the $\MSbar$ scheme in \cite{26,27,28,25,29}. At three
loops fixed moment results are available for moments up to $n$~$=$~$8$ for
$\MSbar$ and $n$~$=$~$7$ for RI$^\prime$, \cite{19,20,24}. The specific Green's
functions relevant for the lattice matching are 
\begin{eqnarray} 
G^{\nu_1 \ldots \nu_n}_W (p) &=& \langle \psi(-p) ~ 
{\cal O}_W^{\nu_1 \ldots \nu_n}(0) ~ \bar{\psi}(p) \rangle \nonumber \\ 
G^{\mu \nu_1 \ldots \nu_n}_T (p) &=& \langle \psi(-p) ~ 
{\cal O}_T^{\mu \nu_1 \ldots \nu_n}(0) ~ \bar{\psi}(p) \rangle 
\label{gfdef}
\end{eqnarray} 
where the operator is inserted at zero momentum. This allows for the
application of the {\sc Mincer} algorithm, \cite{8,9}, which determines the 
finite part of scalar massless two point functions using dimensional 
regularization in $d$~$=$~$4$~$-$~$2\epsilon$ dimensions to three loops. Unlike
earlier approaches to extract anomalous dimensions we do not contract the free 
Lorentz indices with a null vector. Whilst such a contraction has the effect of
excluding the trace terms in the operator itself or the Green's function 
decomposition, the main reason why we cannot follow that route here is that the
lattice makes measurements in different directions of the momentum components. 
This allows for the extraction of the values of each of the individual 
amplitudes into which the Green's functions are decomposed. 

From the form of the operators there will be $n$ $n$-point Feynman rules for 
both the Wilson and transversity operators, each with two quark legs. However, 
at the three loop order we will work at, only the Feynman rules up to and 
including three gluon legs will be necessary. Hence, there are $3$ one loop, 
$37$ two loop and $684$ three loop Feynman diagrams contributing to each 
Green's function. These are generated electronically using the {\sc Qgraf} 
package, \cite{30}, before being converted into {\sc Form} input notation to 
allow for the application of the {\sc Form} version of the {\sc Mincer} 
package, \cite{8,9}. As the {\sc Mincer} algorithm is only applicable to scalar
Feynman integrals, for each moment $n$ the Green's functions, (\ref{gfdef}), 
need to be decomposed into invariant amplitudes and Lorentz tensors which 
respect all symmetry structures. Whilst we will discuss more details later, it 
suffices to note at this point that for Wilson operators there will be two 
independent amplitudes but three for the transversity case.

Although the lattice computations are ultimately performed in the Landau gauge,
we will compute in an arbitrary linear covariant gauge. The associated gauge 
parameter will act as an internal checking parameter since, for instance, in 
the $\MSbar$ scheme the anomalous dimension of gauge invariant operators are
independent of the gauge parameter, \cite{31,32}. As the computations are 
clearly automatic we employ the procedure of \cite{33} where all the diagrams 
are computed for bare coupling, $g_{\mbox{\footnotesize{o}}}$, and gauge 
parameter $\alpha_{\mbox{\footnotesize{o}}}$. To extract the anomalous 
dimension the renormalization constant is fixed (either in the $\MSbar$ or 
RI$^\prime$ schemes) by rescaling these variables by the known coupling 
constant and gauge parameter renormalization constants 
\begin{equation}  
g_{\mbox{\footnotesize{o}}} ~=~ Z_g g ~~~,~~~ 
\alpha_{\mbox{\footnotesize{o}}} ~=~ Z^{-1}_\alpha Z_A \alpha 
\end{equation}
in our conventions, where $Z_A$ is the gluon wave function renormalization 
constant. The remaining divergence for each of the Green's function is absorbed
into the operator renormalization constant together with a specified finite 
part in the case of the RI$^\prime$ scheme to leave the finite parts for the 
lattice matching. In practice the results are determined in the $\MSbar$ scheme
first, primarily due to more consistency checks being available before 
extracting the RI$^\prime$ expressions.  

\sect{RI$^\prime$ scheme.} 

We briefly review the parts of the RI$^\prime$ scheme needed for the 
computations discussed here. Originally the scheme was invented in the context
of the lattice, \cite{21,22}, but it is not restricted to a discrete spacetime.
The continuum analogue has been studied to three and four loop order in 
\cite{23,18}. In general terms it is a mass dependent renormalization scheme 
where the renormalization of the quark field is chosen to be non-minimal in a 
way which is appropriate for lattice analyses. The coupling constant (and thus 
vertex) renormalization is performed in an $\MSbar$ way and so in some sense 
the RI set of schemes sits between $\MSbar$ and MOM type schemes. To reduce 
time (and cost), since taking a derivative on the lattice requires significant 
computation, the quark $2$-point function, $\Sigma_\psi(p)$, is renormalized 
according to the RI$^\prime$ prescription, \cite{21,22}, 
\begin{equation}
\left. \lim_{\epsilon \rightarrow 0} \left[ Z^\prime_\psi \Sigma_\psi(p) 
\right] \right|_{p^2 \, = \, \mu^2} ~=~ \pslash 
\label{ripdef}
\end{equation}
where $\mu^2$ is the renormalization point. In other words there are no
$O(g^2)$ corrections to $\Sigma_\psi(p)$ after renormalization at 
$p^2$~$=$~$\mu^2$ as these finite parts are absorbed into the quark wave
function renormalization constant, $Z^\prime_\psi$. We use the notation
throughout that a ${}^\prime$ on a quantity indicates that renormalization has
been performed in the RI$^\prime$ scheme. Otherwise the scheme is $\MSbar$. For
completeness the RI scheme, which is not of interest to us here, involves
taking a momentum derivative of $\Sigma_\psi(p)$ first before choosing the
result to be the tree value at the renormalization point, \cite{21,22}. As an
extension of the RI$^\prime$ scheme in the continuum, the gluon and 
Faddeev-Popov $2$-point functions are renormalized analogous to (\ref{ripdef}).
However, as most interest in general is in quark $2$-point Green's functions,
there is no real need to pursue this route, unless one was perhaps intending to
consider supersymmetric theories. Similar to the lattice we are ultimately 
interested in converting results from RI$^\prime$ to $\MSbar$ and therefore the
variables in each scheme need to be related. Using the standard conversion
definitions  
\begin{equation}
\alpha^\prime ~=~ \frac{Z^\prime_A}{Z_A} \alpha ~~~,~~~ 
a^\prime ~=~ \frac{Z^\prime_g}{Z_g} a 
\end{equation}
where $a$~$=$~$g^2/(16\pi^2)$, we have the one loop relations, \cite{23,18},
\begin{eqnarray}
a^\prime &=& a ~+~ O( a^5 ) \nonumber \\  
\alpha^\prime 
&=& \left[ 1 + \left( \left( - 9 \alpha^2 - 18 \alpha - 97 \right) C_A 
+ 80 T_F \Nf \right) \frac{a}{36} \right] \alpha ~+~ O(a^2) ~.  
\end{eqnarray} 
The explicit expressions to three loops are available in \cite{18}. Though it 
is worth noting that the Landau gauge is preserved in changing between 
RI$^\prime$ and $\MSbar$. To illustrate the effect the schemes have on the 
basic anomalous dimensions, we note 
\begin{eqnarray} 
\gamma_\psi(a) &=& \alpha C_F a 
+ \frac{1}{4} \left[ ( \alpha^2 + 8 \alpha + 25 ) C_A C_F - 6 C_F^2 
- 8 C_F T_F \Nf \right] a^2 \nonumber \\
\gamma^\prime_\psi(a) &=&  \alpha C_F a ~+~ \left[ \left( 9 \alpha^3 
+ 45 \alpha^2 + 223 \alpha + 225 \right) C_A \right. \nonumber \\
&& \left. ~~~~~~~~~~~~~~~~~~~~-~ 54 C_F - \left( 80 \alpha + 72 \right) T_F \Nf 
\right] \frac{C_F a^2}{36} 
\label{rirelns}
\end{eqnarray} 
where the group theoretic quantities are defined by  
\begin{equation} 
\mbox{Tr} \left( T^a T^b \right) ~=~ T_F \delta^{ab} ~~~,~~~ 
T^a T^a ~=~ C_F I ~~~,~~~ f^{acd} f^{bcd} ~=~ C_A \delta^{ab}
\end{equation} 
for a colour group with generators $T^a$ and structure functions $f^{abc}$. 
Clearly the difference in the numerical structure in (\ref{rirelns}) only 
appears at two and higher loops. 

For the flavour non-singlet operators we are interested in here, we follow a
similar route to (\ref{ripdef}) for defining the operator renormalization
constant in the RI$^\prime$ scheme. Writing $\Sigma^{(T)}_{{\cal O}}(p)$ as the
amplitude in the Lorentz decomposition of (\ref{gfdef}) which contains the
tree, $(T)$, part of the operator, we set  
\begin{equation}
\left. \lim_{\epsilon \, \rightarrow \, 0} \left[ Z^\prime_\psi  
Z^\prime_{{\cal O}} \Sigma^{(T)}_{{\cal O}}(p) \right] \right|_{p^2 \, = \, 
\mu^2} ~=~ {\cal T}
\end{equation}
where ${\cal T}$ is the value of the tree term of amplitude, which may not 
necessarily be unity given the specific (non-unique) way of carrying out the
decomposition. 

\sect{Simple example.} 

We now illustrate the preceeding remarks by discussing the case of the 
$n$~$=$~$2$ transversity operator in more detail, \cite{19}. First, given the 
symmetry properties (\ref{trasymm}) the explicit traceless symmetrized operator
is 
\begin{eqnarray} 
{\cal S} \bar{\psi} \sigma^{\mu\nu} D^\rho \psi &=& \bar{\psi} \sigma^{\mu\nu} 
D^\rho \psi ~+~ \bar{\psi} \sigma^{\mu\rho} D^\nu \psi ~-~ \frac{2}{(d-1)} 
\eta^{\nu\rho} \bar{\psi} \sigma^{\mu\lambda} D_\lambda \psi \nonumber \\
&& +~ \frac{1}{(d-1)} \left( \eta^{\mu\nu} \bar{\psi} \sigma^{\rho\lambda} 
D_\lambda \psi + \eta^{\mu\rho} \bar{\psi} \sigma^{\nu\lambda} D_\lambda \psi 
\right) 
\label{tra2def} 
\end{eqnarray} 
in $d$-dimensions. Inserting (\ref{tra2def}) into the Green's function
$G^{\mu\nu\rho}_T(p)$, it is decomposed into the three invariant amplitudes as 
\begin{eqnarray} 
G^{\mu\nu\rho}_T(p) &=& \Sigma^{(1)}_T(p) 
\left( \sigma^{\mu\nu} p^\rho + \sigma^{\mu\rho} p^\nu - \frac{(d+2)}{p^2} 
\sigma^{\mu\lambda} p^\nu p^\rho p_\lambda + \eta^{\nu\rho} 
\sigma^{\mu\lambda} p_\lambda \right) \nonumber \\
&& +~ \Sigma^{(2)}_T(p) 
\left( \eta^{\mu\nu} \sigma^{\rho\lambda} p_\lambda + \eta^{\mu\rho} 
\sigma^{\nu\lambda} p_\lambda - (d+1) \eta^{\nu\rho} \sigma^{\mu\lambda} 
p_\lambda \right. \nonumber \\
&& \left. ~~~~~~~~~~~~~~~~~~~~+~ \frac{(d-1)(d+2)}{p^2} \sigma^{\mu\lambda} 
p^\nu p^\rho p_\lambda \right) \nonumber \\ 
&& +~ \Sigma^{(3)}_T(p) 
\left( \sigma^{\nu\lambda} p^\mu p^\rho p_\lambda + \sigma^{\rho\lambda} p^\mu 
p^\nu p_\lambda + d \sigma^{\mu\lambda} p^\nu p^\rho p_\lambda 
- \eta^{\nu\rho} \sigma^{\mu\lambda} p_\lambda p^2 \right) 
\end{eqnarray} 
in $d$-dimensions. It is worth noting that this and other decompositions are 
not unique since one can always take a linear combination of the three
(independent) tensor structures consistent with the symmetry and traceless
properties to form another set of independent amplitudes. However, with this
choice one can algebraically form a scalar object which is computed via
{\sc Mincer}. For instance, \cite{19},  
\begin{eqnarray} 
\Sigma^{(1)}_T(p) &=& 
-~ \frac{1}{8(d-1)(d-2)} \mbox{tr} \left[ \left( \sigma_{\mu\nu} p_\rho
+ \sigma_{\mu\rho} p_\nu - \frac{(d+2)}{p^2} \sigma_{\mu\lambda} p_\nu p_\rho 
p^\lambda \right. \right. \nonumber \\
&& \left. \left. ~~~~~~~~~~~~~~~~~~~~~~~~~~~~+~ \eta_{\nu\rho} 
\sigma_{\mu\lambda} p^\lambda \right) G^{\mu\nu\rho}_T(p) \right] \nonumber \\ 
&& -~ \frac{1}{8(d-1)(d-2)p^2} \mbox{tr} \left[ \left( \sigma_{\nu\lambda} 
p_\mu p_\rho p^\lambda + \sigma_{\rho\lambda} p_\mu p_\nu p^\lambda \right. 
\right.  \nonumber \\ 
&& \left. \left. ~~~~~~~~~~~~~~~~~~~~~~~~~~~~~~+~ d \sigma_{\mu\lambda} p_\nu 
p_\rho p^\lambda - \eta_{\nu\rho} \sigma_{\mu\lambda} p^\lambda p^2 \right) 
G^{\mu\nu\rho}_T (p) \right] ~.  
\end{eqnarray}
This together with $\Sigma^{(i)}_T(p)$, $i$~$=$~$2$ and $3$, are the objects
of interest for the lattice matching and have been determined to $O(a^3)$,
\cite{19}. For this specific example, we note that the construction of the
tensor basis as well as the amplitude decomposition can easily be carried out
by hand. This is primarily due to the small number of free Lorentz indices
present. Clearly for the extraction of the anomalous dimensions and amplitudes
for the higher moment operators, such a procedure would be unacceptably time
consuming by hand. Moreover, it would be prone to elementary algebraic errors.  

\sect{Higher moment issues.} 

To extract the anomalous dimensions for the higher moment operators, it is
clear that one has to proceed with a computer algebra construction to determine
the basis for the independent amplitudes and hence the projections. We discuss 
the issues in relation to the $n$~$=$~$8$ transversity operator as an example,
\cite{24}. For this case there are initially seventeen potential tensors into
which the Green's function (\ref{gfdef}) can be decomposed. These are built
from the relevant vectors and tensors of the operator in question, which for
transversity are $p^\mu$, $\eta^{\mu\nu}$ and $\sigma^{\mu\nu}$. The only
constraint being that the Green's function has nine independent indices. Given 
these seventeen tensors then within {\sc Form} it is straightforward to 
construct the seventeen tensors which have the correct symmetry, but {\em not} 
traceless, properties. Taking a linear combination of these new objects with as
yet unrelated coefficients, the relationship between these are fixed by 
imposing the remaining traceless criterion. In practical terms we take 
successive pairs of free indices and contract them. The coefficients of the 
resulting tensors produce constraints on the seventeen initial coefficients 
which can be solved. Whilst there are more contractions than coefficients there
is redundancy in the system of linear equations which determine the 
coefficients. This is due to the symmetry of the operator itself. However, 
there is no unique solution and three coefficients remain unrelated producing 
three independent amplitudes. (For the Wilson operator the corresponding number
is two.) For the higher moments, as we are ultimately interested in the 
anomalous dimensions, the specific linear combination one uses is not a major 
issue. The only constraint is to choose that projector of the three which leads
to the lowest computation time when {\sc Mincer} is applied. The test for this 
is to compare the run times for each projector to do the full {\em two} loop 
calculation in an arbitrary linear covariant gauge before generating the 
results for the three loop diagrams.

The other main computer algebra issue is the construction of the Feynman rules
for each operator. Given that the Green's function has free indices one in
principle has to construct the full symmetrized and traceless operator before
applying the {\sc Form} routine to generate the explicit Feynman rules for the
operator. However, given that the Green's function will be multiplied by a 
projector which is traceless, that part of the operator containing
$\eta_{\mu\nu}$ tensors will automatically give zero upon contracting with the
projector. Therefore there is no need to have an operator which is traceless;
only an operator which is symmetrized will suffice. This will reduce 
computation time since otherwise with a traceless operator there will be an
internal intermediate expression swell which will be sorted by {\sc Form} to
produce the equivalent scalar expression as ignoring the traceless part. For
instance, for the $n$~$=$~$8$ transversity operator the expression swell would 
have been substantial. Finally, in relation to the Feynman rules, only the part
of the operator up to and including two quark and three gluon leg insertions 
are required for the full three loop computation. To illustrate the size of our
higher moment calculation the {\sc Form} module containing the operator Feynman
rule was $36$ Mbytes for $n$~$=$~$7$ transversity and $300$ Mbytes for 
$n$~$=$~$8$ transversity, \cite{24}. Indeed the latter calculation could only 
be performed in the Feynman gauge rather than the full linear covariant or 
Landau gauges. Even then it took of the order of $40$ days on a dual opteron 
$64$ bit SMP $2$GHz machine. Hence, only the $\MSbar$ result was determined 
with the RI$^\prime$ scheme anomalous dimension yet to be established. 

As with all large computations carried out symbolically, it is worth detailing
the various checks we used in order to be confident that our results are
credible. First, for the case of the Wilson operators the three loop $\MSbar$
anomalous dimensions are known, \cite{1,2,3,4}, and our anomalous dimensions 
must therefore agree before extracting any finite parts for lattice matching.
Moreover, for both Wilson and transversity operators the $\MSbar$ expressions
have been shown to be independent of the linear covariant gauge fixing
parameter. For the transversity case we have the checks that the two loop
anomalous dimensions must agree with \cite{26,27,28,25,29} for the various $n$ 
we consider. At three loops the only substantial check is that the residues of 
the poles in $1/\epsilon^2$ and $1/\epsilon^3$ have to agree with the 
renormalization group consistency check. In other words these are predicted 
from the one and two loop parts of the anomalous dimensions. In addition for 
the RI$^\prime$ scheme, one can compute the anomalous dimensions either 
directly from the renormalization constants deduced from the Green's function 
itself, or from the conversion functions, $C_{\cal O}(a,\alpha)$, based on the 
renormalization group. This is defined by 
\begin{equation}
C_{\cal O}(a,\alpha) ~=~ \frac{Z^\prime_{\cal O}}{Z_{\cal O}} 
\end{equation}
where the renormalization constants are both expressed in terms of the 
variables defined in the {\em same} scheme. Then the RI$^\prime$ anomalous
dimension is given by 
\begin{equation}
\gamma^\prime_{\cal O} \left(a^\prime\right) ~=~ 
\gamma_{\cal O} (a) ~-~ \beta(a) \frac{\partial ~}{\partial a} 
\ln C_{\cal O} (a,\alpha) ~-~ \alpha \gamma_\alpha (a) 
\frac{\partial ~}{\partial \alpha} \ln C_{\cal O} (a,\alpha) ~.  
\end{equation}
(See, for example, \cite{34}.) Therefore, the expression on the left side will 
agree with the direct renormalization. For all the results presented in the 
next section we note that they all pass the checks discussed here.  

\sect{Results.} 

First, we record the explicit values for the anomalous dimensions of the Wilson
operators $n$~$=$~$5$ and $6$ in RI$^\prime$ for arbitrary $\alpha$, which are 
new. The notation is that the numerical subscript denotes the moment whilst the
superscript, $W$ or $T$, corresponds to either the Wilson or transversity 
operator respectively. We find  
\begin{eqnarray} 
\gamma^{\prime \, W}_5(a) &=& \frac{91}{15} C_F a ~+~ \left[ 
\left( 33525 \alpha^2 + 100575 \alpha + 1729270 \right) C_A \right. 
\nonumber \\
&& \left. ~~~~~~~~~~~~~~~~~~~~~~~-~ 156114 C_F - 673880 T_F \Nf \right] 
\frac{C_F a^2}{27000} \nonumber \\ 
&& +~ \left[ \left( 30172500 \alpha^4 + 289359000 \alpha^3 
- 97200000 \zeta_3 \alpha^2 + 1409428125 \alpha^2 \right. \right. \nonumber \\
&& \left. \left. ~~~~~~-~ 1004400000 \zeta_3 \alpha + 4758071625 \alpha 
- 5142528000 \zeta_3 + 52067172425 \right) C_A^2 \right. \nonumber \\
&& \left. ~~~~~+~ \left( 23726250 \alpha^3 + 30956400 \alpha^2 
- 415630950 \alpha \right. \right. \nonumber \\
&& \left. \left. ~~~~~~~~~~~~+~ 102384000 \zeta_3 - 9145680720 \right) 
C_A C_F ~+~ 6023484800 T_F^2 \Nf^2 \right. \nonumber \\ 
&& \left. ~~~~~-~ \left( 268200000 \alpha^2 - 259200000 \zeta_3 \alpha 
+ 1582173000 \alpha \right. \right. \nonumber \\ 
&& \left. \left. ~~~~~~~~~~~~+~ 2514240000 \zeta_3 + 36792205400 \right) 
C_A T_F \Nf \right. \nonumber \\
&& \left. ~~~~~+~ \left( 107259600 \alpha + 3680640000 \zeta_3 
- 3053173120 \right) C_F T_F \Nf \right. \nonumber \\
&& \left. ~~~~~+~ \left( 1832544000 \zeta_3 - 829297168 \right) C_F^2 
\right] \frac{C_F a^3}{48600000} ~+~ O(a^4) 
\end{eqnarray}  
and 
\begin{eqnarray} 
\gamma^{\prime \, W}_6(a) &=& \frac{709}{105} C_F a ~+~ \left[ 
\left( 12116475 \alpha^2 + 36349425 \alpha + 670295290 \right) C_A \right. 
\nonumber \\
&& \left. ~~~~~~~~~~~~~~~~~~~~~~~-~ 57119598 C_F - 263443880 T_F \Nf \right] 
\frac{C_F a^2}{9261000} \nonumber \\ 
&& +~ \left[ \left( 534336547500 \alpha^4 + 5228103069000 \alpha^3 
- 1750329000000 \zeta_3 \alpha^2 \right. \right. \nonumber \\
&& \left. \left. ~~~~~~+~ 25439835416625 \alpha^2 
- 18086733000000 \zeta_3 \alpha + 86004002776125 \alpha \right. \right. 
\nonumber \\
&& \left. \left. ~~~~~~-~ 92121315552000 \zeta_3 
+ 988839358918775 \right) C_A^2 \right. \nonumber \\
&& \left. ~~~~~+~ \left( 355203339750 \alpha^3 + 158333464800 \alpha^2 
- 9375191062650 \alpha \right. \right. \nonumber \\
&& \left. \left. ~~~~~~~~~~~~-~ 5509035504000 \zeta_3 
- 172078530172080 \right) C_A C_F \right. \nonumber \\
&& \left. ~~~~~-~ \left( 4749658200000 \alpha^2 - 4667544000000 \zeta_3 \alpha 
+ 28468726629000 \alpha \right. \right. \nonumber \\ 
&& \left. \left. ~~~~~~~~~~~~+~ 49102562880000 \zeta_3 + 704961641573000 
\right) C_A T_F \Nf \right. \nonumber \\
&& \left. ~~~~~+~ \left( 2419404145200 \alpha + 73311557760000 \zeta_3 
- 59288998908160 \right) C_F T_F \Nf \right. \nonumber \\
&& \left. ~~~~~+~ \left( 31055615136000 \zeta_3 - 13674447985168 \right) C_F^2 
\right. \nonumber \\ 
&& \left. ~~~~~+~ 117065906115200 T_F^2 \Nf^2 
\right] \frac{C_F a^3}{816820200000} ~+~ O(a^4) 
\end{eqnarray}  
where $\zeta_n$ is the Riemann zeta function. As the Landau gauge is of 
particular interest, we record that the previous two expressions when 
$\alpha$~$=$~$0$ are  
\begin{eqnarray} 
\left. \gamma^{\prime \, W}_{5}(a) \right|_{\alpha = 0} &=& 
\frac{91C_F}{15} a ~+~ [ 864635 C_A - 78057 C_F - 336940 T_F \Nf ] 
\frac{C_F a^2}{13500} \nonumber \\ 
&& +~ \left[ ( 52067172425 - 5142528000 \zeta_3 ) C_A^2 \right. \nonumber \\
&& \left. ~~~~~+~ ( 102384000 \zeta_3 - 9145680720 ) C_A C_F \right. 
\nonumber \\
&& \left. ~~~~~-~ ( 2514240000 \zeta_3 + 36792205400 ) C_A T_F \Nf \right. 
\nonumber \\
&& \left. ~~~~~+~ ( 1832544000 \zeta_3 - 829297168 ) C_F^2 \right. 
\nonumber \\
&& \left. ~~~~~+~ ( 3680640000 \zeta_3 - 3053173120 ) C_F T_F \Nf \right. 
\nonumber \\
&& \left. ~~~~~+~ 6023484800 T_F^2 \Nf^2 \right] \frac{C_F a^3}{48600000} ~+~ 
O(a^4)  
\end{eqnarray}  
and 
\begin{eqnarray} 
\left. \gamma^{\prime \, W}_{6}(a) \right|_{\alpha = 0} &=& 
\frac{709C_F}{105} a \nonumber \\
&& +~ [ 335147645 C_A - 28559799 C_F - 131721940 T_F \Nf ] 
\frac{C_F a^2}{4630500} \nonumber \\ 
&& +~ \left[ ( 988839358918775 - 92121315552000 \zeta_3 ) C_A^2 \right. 
\nonumber \\
&& \left. ~~~~~-~ ( 5509035504000 \zeta_3 - 172078530172080 ) C_A C_F \right. 
\nonumber \\
&& \left. ~~~~~-~ ( 49102562880000 \zeta_3 + 704961641573000 ) C_A T_F \Nf 
\right. \nonumber \\
&& \left. ~~~~~+~ ( 31055615136000 \zeta_3 - 13674447985168 ) C_F^2 \right. 
\nonumber \\
&& \left. ~~~~~+~ ( 73311557760000 \zeta_3 - 59288998908160 ) C_F T_F \Nf 
\right. \nonumber \\
&& \left. ~~~~~+~ 117065906115200 T_F^2 \Nf^2 \right] 
\frac{C_F a^3}{816820200000} ~+~ O(a^4) ~.  
\end{eqnarray}  
For further comparison between the schemes the $\MSbar$ and RI$^\prime$ 
expressions for $n$~$=$~$5$ transversity are, \cite{20}, 
\begin{eqnarray}  
\gamma^T_5(a) &=& \frac{92}{15} C_F a ~+~ \left[ 189515 C_A 
- 41674 C_F - 79810 T_F \Nf \right] \frac{C_F a^2}{6750} \nonumber \\ 
&& +~ \left[ \left( 190836000 \zeta_3 + 1975309075 \right) C_A^2 \right.
\nonumber \\
&& \left. ~~~~~-~ \left( 572508000 \zeta_3 + 325464235 \right) C_A C_F 
\right. \nonumber \\
&& \left. ~~~~~-~ \left( 1192320000 \zeta_3 + 511395100 \right) C_A T_F \Nf
\right. \nonumber \\
&& \left. ~~~~~+~ \left( 381672000 \zeta_3 - 254723696 \right) C_F^2 
\right. \nonumber \\
&& \left. ~~~~~+~ \left( 1192320000 \zeta_3 - 989903260 \right) C_F T_F \Nf 
\right. \nonumber \\
&& \left. ~~~~~-~ 83718800 T_F^2 \Nf^2 \right] \frac{C_F a^3}{12150000} ~+~ 
O(a^4) 
\end{eqnarray}  
and 
\begin{eqnarray} 
\gamma^{\prime \, T}_5(a) &=& \frac{92}{15} C_F a ~+~ \left[ 
\left( 30825 \alpha^2 + 92475 \alpha + 1740690 \right) C_A \right. 
\nonumber \\
&& \left. ~~~~~~~~~~~~~~~~~~~~~~~-~ 166696 C_F - 676560 T_F \Nf \right] 
\frac{C_F a^2}{27000} \nonumber \\ 
&& +~ \left[ \left( 194197500 \alpha^4 + 1854279000 \alpha^3 
- 583200000 \zeta_3 \alpha^2 + 8993896875 \alpha^2 \right. \right. 
\nonumber \\
&& \left. \left. ~~~~~~-~ 6026400000 \zeta_3 \alpha + 30074295375 \alpha 
- 37353312000 \zeta_3 \right. \right. \nonumber \\
&& \left. \left. ~~~~~~+~ 356401468700 \right) C_A^2 
~+~ \left( 91239750 \alpha^3 - 209956950 \alpha^2 \right. \right. \nonumber \\
&& \left. \left. ~~~~~~~~~~~~~~~~-~ 4997987400 \alpha + 1076976000 \zeta_3 
- 60979980560 \right) C_A C_F \right. \nonumber \\
&& \left. ~~~~~-~ \left( 1726200000 \alpha^2 - 1555200000 \zeta_3 \alpha 
+ 10041363000 \alpha \right. \right. \nonumber \\ 
&& \left. \left. ~~~~~~~~~~~~+~ 17858880000 \zeta_3 + 253330505600 \right) 
C_A T_F \Nf \right. \nonumber \\
&& \left. ~~~~~+~ 41629683200 T_F^2 \Nf^2 \right. \nonumber \\ 
&& \left. ~~~~~+~ \left( 1289803200 \alpha + 27164160000 \zeta_3 
- 22363266560 \right) C_F T_F \Nf \right. \nonumber \\
&& \left. ~~~~~+~ \left( 10686816000 \zeta_3 - 7132263488 \right) C_F^2 
\right] \frac{C_F a^3}{340200000} ~+~ O(a^4) ~.  
\end{eqnarray}  
For completeness we record the next $\MSbar$ anomalous dimensions in the 
sequence are, \cite{20}, 
\begin{eqnarray}  
\gamma^T_6(a) &=& \frac{34}{5} C_F a ~+~ \left[ 204770 C_A 
- 42129 C_F - 88810 T_F \Nf \right] \frac{C_F a^2}{6750} \nonumber \\ 
&& +~ \left[ \left( 707616000 \zeta_3 + 7527909825 \right) C_A^2 
\right. \nonumber \\
&& \left. ~~~~~-~ \left( 2122848000 \zeta_3 + 1373507730 \right) C_A C_F 
\right. \nonumber \\
&& \left. ~~~~~-~ \left( 4626720000 \zeta_3 + 1841332000 \right) C_A T_F \Nf 
\right. \nonumber \\
&& \left. ~~~~~+~ \left( 1415232000 \zeta_3 - 684744816 \right) C_F^2 
\right. \nonumber \\
&& \left. ~~~~~+~ \left( 4626720000 \zeta_3 - 3910683210 \right) C_F T_F \Nf 
\right. \nonumber \\
&& \left. ~~~~~-~ 320975800 T_F^2 \Nf^2 \right]
\frac{C_F a^3}{42525000} ~+~ O(a^4) 
\end{eqnarray}  
and 
\begin{eqnarray}  
\gamma^T_7(a) &=& \frac{258}{35} C_F a ~+~ \left[ 75266555 C_A 
- 15484767 C_F - 33149830 T_F \Nf \right] \frac{C_F a^2}{2315250} \nonumber \\ 
&& +~ \left[ \left( 3517994592000 \zeta_3 + 38365845513450 \right) C_A^2 
\right. \nonumber \\
&& \left. ~~~~~-~ \left( 10553983776000 \zeta_3 + 5978407701105 \right)
C_A C_F \right. \nonumber \\
&& \left. ~~~~~-~ \left( 24084527040000 \zeta_3 + 9039144860900 \right) 
C_A T_F \Nf \right. \nonumber \\
&& \left. ~~~~~+~ \left( 7035989184000 \zeta_3 - 4192441946262 \right) C_F^2 
\right. \nonumber \\
&& \left. ~~~~~+~ \left( 24084527040000 \zeta_3 - 20698675427220 \right)
C_F T_F \Nf \right. \nonumber \\
&& \left. ~~~~~-~ 1651311191600 T_F^2 \Nf^2 \right]
\frac{C_F a^3}{204205050000} ~+~ O(a^4) ~.  
\end{eqnarray}  
The complete set of three loop transversity anomalous dimensions in $\MSbar$
and RI$^\prime$ are given in \cite{18,19,20,24}. 

\sect{Conclusions.} 

We conclude with a few brief remarks. First, the three loop anomalous 
dimensions are available for the transversity operator for each moment up to
$n$~$=$~$8$ in the $\MSbar$ scheme and $n$~$=$~$7$ for the lattice motivated
RI$^\prime$ scheme. The former in particular will provide important independent
checks for future explicit arbitrary moment evaluations of the three loop 
anomalous dimensions. A by-product of the overall project, \cite{18,19,20}, has
been the provision of the finite parts of a Green's function which are 
necessary for lattice measurements of matrix elements. The three loop 
perturbative information is essential to obtaining more precise numerical 
estimates. In addition we have given the (new) RI$^\prime$ anomalous dimensions
for the $n$~$=$~$5$ and $6$ Wilson operators at three loops. Whilst it is in 
principle possible to continue with the computation of the transversity higher 
moments to $n$~$=$~$9$ and beyond, the present method has become too tedious. 
This is primarily due to the increase in the number of free Lorentz indices on 
the operator which was originally required for the lattice comparison. 
Moreover, the actual computation time as indicated for $n$~$=$~$8$ in the 
Feynman gauge has already become unacceptably long. An explicit arbitrary $n$
calculation exploiting the algorithm of \cite{1,2,3,4} would achieve all moment
information during one run, at possibly a computation time which is not too 
dissimilar from that for one high moment.  

\vspace{1cm}
\noindent
{\bf Acknowledgements.} The author thanks Dr P.E.L. Rakow and Dr C. McNeile
for valuable discussions.


\begin{thebibliography}{99} 
\bibitem{1} S. Moch, J.A.M. Vermaseren \& A. Vogt, Nucl. Phys. {\bf B688}
(2004), 101. 
\bibitem{2} S. Moch, J.A.M. Vermaseren \& A. Vogt, Nucl. Phys. {\bf B691}
(2004), 129. 
\bibitem{3} S. Moch, J.A.M. Vermaseren \& A. Vogt, Nucl. Phys. Proc. Suppl.
{\bf 135} (2004), 137. 
\bibitem{4} S. Moch, J.A.M. Vermaseren \& A. Vogt, Phys. Lett. {\bf B606}
(2005), 123. 
\bibitem{5} J.A.M. Vermaseren, math-ph/0010025. 
\bibitem{6} S.A. Larin, T. van Ritbergen \& J.A.M. Vermaseren, Nucl. Phys. 
{\bf B427} (1994), 41. 
\bibitem{7} S.A. Larin, P. Nogueira, T. van Ritbergen \& J.A.M. Vermaseren, 
Nucl. Phys. {\bf B492} (1997), 338. 
\bibitem{8} S.G. Gorishny, S.A. Larin, L.R. Surguladze \& F.K. Tkachov,
Comput. Phys. Commun. {\bf 55} (1989), 381. 
\bibitem{9} S.A. Larin, F.V. Tkachov \& J.A.M. Vermaseren, ``The Form version
of Mincer'', NIKHEF-H-91-18. 
\bibitem{10} A. R\'{e}tey \& J.A.M. Vermaseren, Nucl. Phys. {\bf B604} (2001),
281. 
\bibitem{11} J. Bl\"{u}mlein \& J.A.M. Vermaseren, Phys. Lett. {\bf B606} 
(2005), 130. 
\bibitem{12} J.P. Ralston \& D.E. Soper, Nucl. Phys. {\bf B152} (1979), 109. 
\bibitem{13} R.L. Jaffe \& X. Ji, Phys. Rev. Lett. {\bf 67} (1991), 552;  
Nucl. Phys. {\bf B375} (1992), 527. 
\bibitem{14} J.I. Cortes, B. Pire \& J.P. Ralston, Z. Phys. {\bf C55} (1992), 
409. 
\bibitem{15} M. G\"{o}ckeler, R. Horsley, H. Oelrich, H. Perlt, D. Petters, 
P.E.L. Rakow, A. Sch\"{a}fer, G. Schierholz \& A. Schiller, Nucl. Phys. 
{\bf B544} (1999), 699.  
\bibitem{16} S. Capitani, M. G\"{o}ckeler, R. Horsley, H. Perlt, P.E.L. Rakow,
G. Schierholz \& A. Schiller, Nucl. Phys. {\bf B593} (2001), 183.   
\bibitem{17} M. G\"{o}ckeler, R. Horsley, D. Pleiter, P.E.L. Rakow, 
A. Sch\"{a}fer and G. Schierholz, Nucl. Phys. Proc. Suppl. {\bf 119} (2003),
23. 
\bibitem{18} J.A. Gracey, Nucl. Phys. {\bf B662} (2003), 247. 
\bibitem{19} J.A. Gracey, Nucl. Phys. {\bf B667} (2003), 242. 
\bibitem{20} J.A. Gracey, JHEP {\bf 0610} (2006), 040.
\bibitem{21} G. Martinelli, C. Pittori, C.T. Sachrajda, M. Testa \& A. 
Vladikas, Nucl. Phys. {\bf B445} (1995), 81. 
\bibitem{22} E. Franco \& V. Lubicz, Nucl. Phys. {\bf B531} (1998), 641
\bibitem{23} K.G. Chetyrkin \& A. R\'{e}tey, Nucl. Phys. {\bf B583} (2000), 3. 
\bibitem{24} J.A. Gracey, Phys. Lett. {\bf B643} (2006), 374. 
\bibitem{25} A. Hayashigaki, Y. Kanazawa \& Y. Koike, Phys. Rev. {\bf D56}
(1997), 7350.  
\bibitem{26} F. Baldracchini, N.S. Craigie, V. Roberto \& M. Socolovsky,
Fortschr. Phys. {\bf 30} (1981), 505. 
\bibitem{27} X. Artru \& M. Mekhfi, Z. Phys. {\bf C45} (1990), 669. 
\bibitem{28} W. Vogelsang, Phys. Rev. {\bf D57} (1998), 1886.  
\bibitem{29} J. Bl\"{u}mlein, Eur. Phys. J. {\bf C20} (2001), 683. 
\bibitem{30} P. Nogueira, J. Comput. Phys. {\bf 105} (1993), 279. 
\bibitem{31} G. 't Hooft, Nucl. Phys. {\bf B61} (1973), 455. 
\bibitem{32} W.E. Caswell \& F. Wilczek, Phys. Lett. {\bf B49} (1974), 291. 
\bibitem{33} S.A. Larin \& J.A.M. Vermaseren, Phys. Lett. {\bf B303} (1993), 
334. 
\bibitem{34} J.C. Collins, {\it Renormalization} (Cambridge University Press,
1984). 
\end{thebibliography}
\end{document}